\documentclass[final,1p,times]{elsarticle} 
\usepackage{graphicx} 
\usepackage{amssymb} 
\usepackage{amsthm} 
\usepackage{lineno} 

\journal{Nuclear Physics A} 
\begin{document} 

\begin{frontmatter} 


\title{Status of ATLAS and Preparation for the Pb-Pb Run}

\author{Ji\v{r}\'{i} Dolej\v{s}\'{i} $^{a}$ for the ATLAS Collaboration}

\address[a]{Charles University, Faculty of Mathematics and Physics, IPNP,\\
V Holesovickach 2,
CZ-180 00 Praha 8, Czech Republic}

\begin{abstract} 
The status and performance of the ATLAS detector will be discussed, with a
view towards the Pb+Pb run expected in 2010. The ATLAS experiment took its first beam data in September 2008 and is actively preparing for the planned start of LHC collision data-taking in 2009.  This preparation includes hardware and software
commissioning, as well as calibration and cosmic data analysis.  
\end{abstract} 

\end{frontmatter} 



\section{Performance of ATLAS and experience from the 2008 run}

This contribution aims to illustrate the ATLAS performance observed during the beam tests, cosmic runs and at the moment of first protons injected into LHC. Refs. \cite{jinst} and \cite{CSC} provide a comprehensive overview of ATLAS and its expected performance.

\begin{figure}[ht]
\centering
\includegraphics[width=0.9\textwidth]{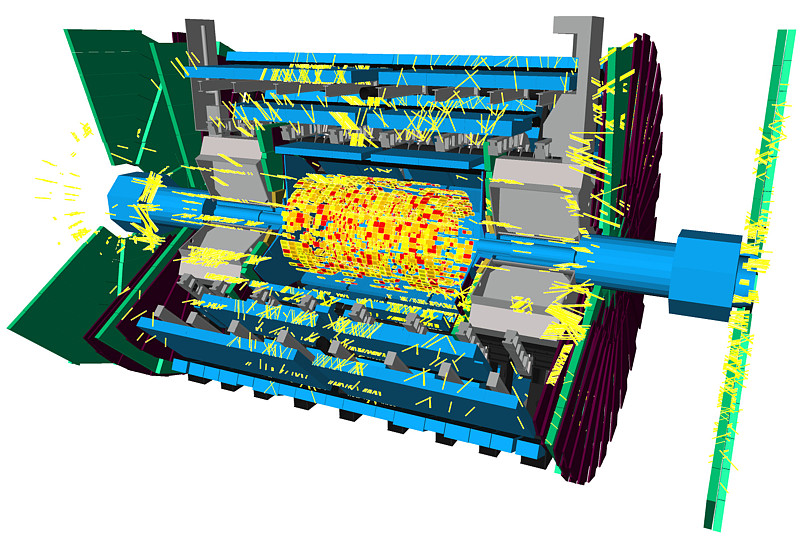}
\caption[]{(Color online) Single beam splash event on September 10th, 2008: particles from a beam hitting the collimator 140 m from ATLAS are detected by the whole detector.}
\label{firstEventVP1}
\end{figure}

\begin{figure}[ht]
\centering
\includegraphics[width=0.6\textwidth]{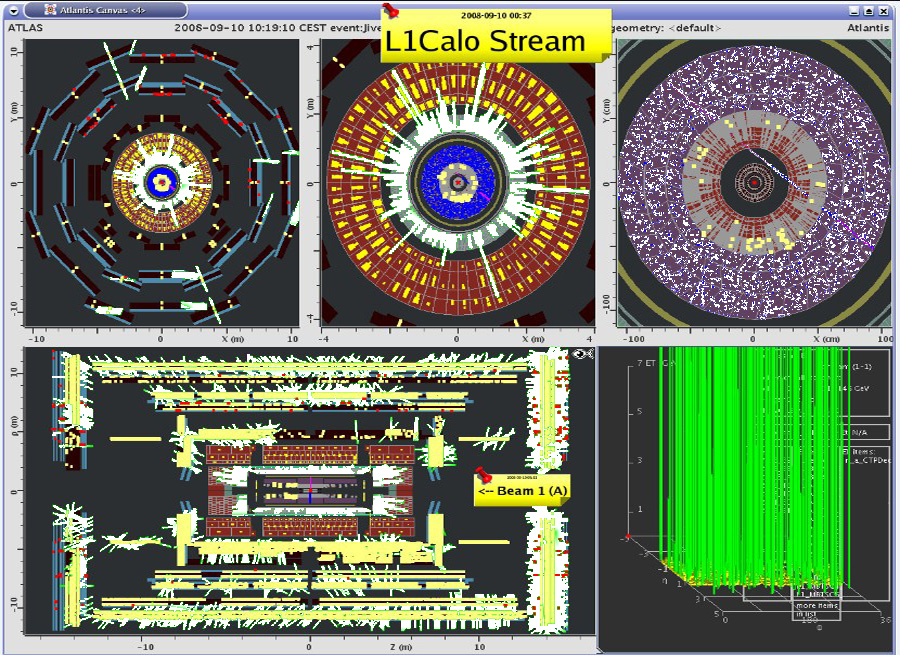}
\caption[]{Single beam splash event on September 10th, 2008: Event display with a more detailed look at muon detectors, calorimeters and the inner detector in the top row.}
\label{lhc_atlas1}
\end{figure}

\subsection*{Inner detector}
Most data come from the cosmic run which started on September 14th, 2008:

\begin{figure}[hbt]
\centering
\includegraphics[width=1.0\textwidth]{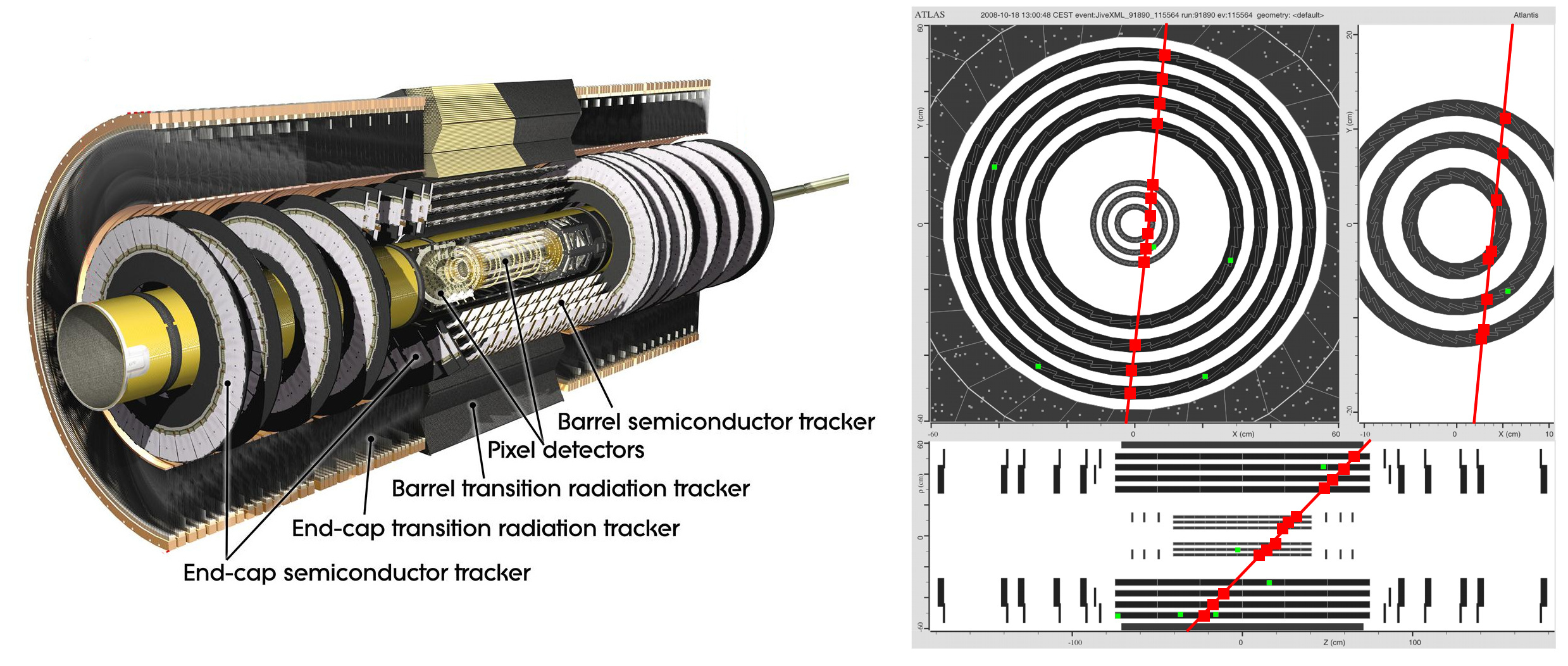}
\caption[]{{\bf Left panel:} The ATLAS inner detector. {\bf Right panel:} Display of cosmic ray event going through the pixel detector. }
\label{InnerD}
\end{figure}

The whole inner detector performed remarkably well -- the pixel detector reached efficiency of about 98.4\%, SCT (SemiConductor Tracker) Barrel and of SCT EndCaps were operational with efficiency of
99\% and 97\% respectively with 99\% of all modules operational and with 99.8\% strips giving signal. 
The TRT (Transition Radiation Tracker) was found to have 97\% channels reading out and ~99\% of electronics working. TRT recorded in total 2.9 M tracks. 
Repairs and improvements, especially of the cooling, are scheduled for 2009.

The right panel of Fig. \ref{InnerD} shows a cosmic ray event going through the SCT and pixel detector. Shown are the XY view (of SCT and pixels and of pixels alone) and an RZ view. The track has a hit in each of the layers in both the upper and the lower hemisphere (with two hits in the innermost pixel layer due to a module overlap). Apart from the signal hits there is only one other hit in the pixel detector demonstrating the very low noise level in the detector.

\begin{figure}[ht]
\centering
\includegraphics[width=0.35\textwidth]{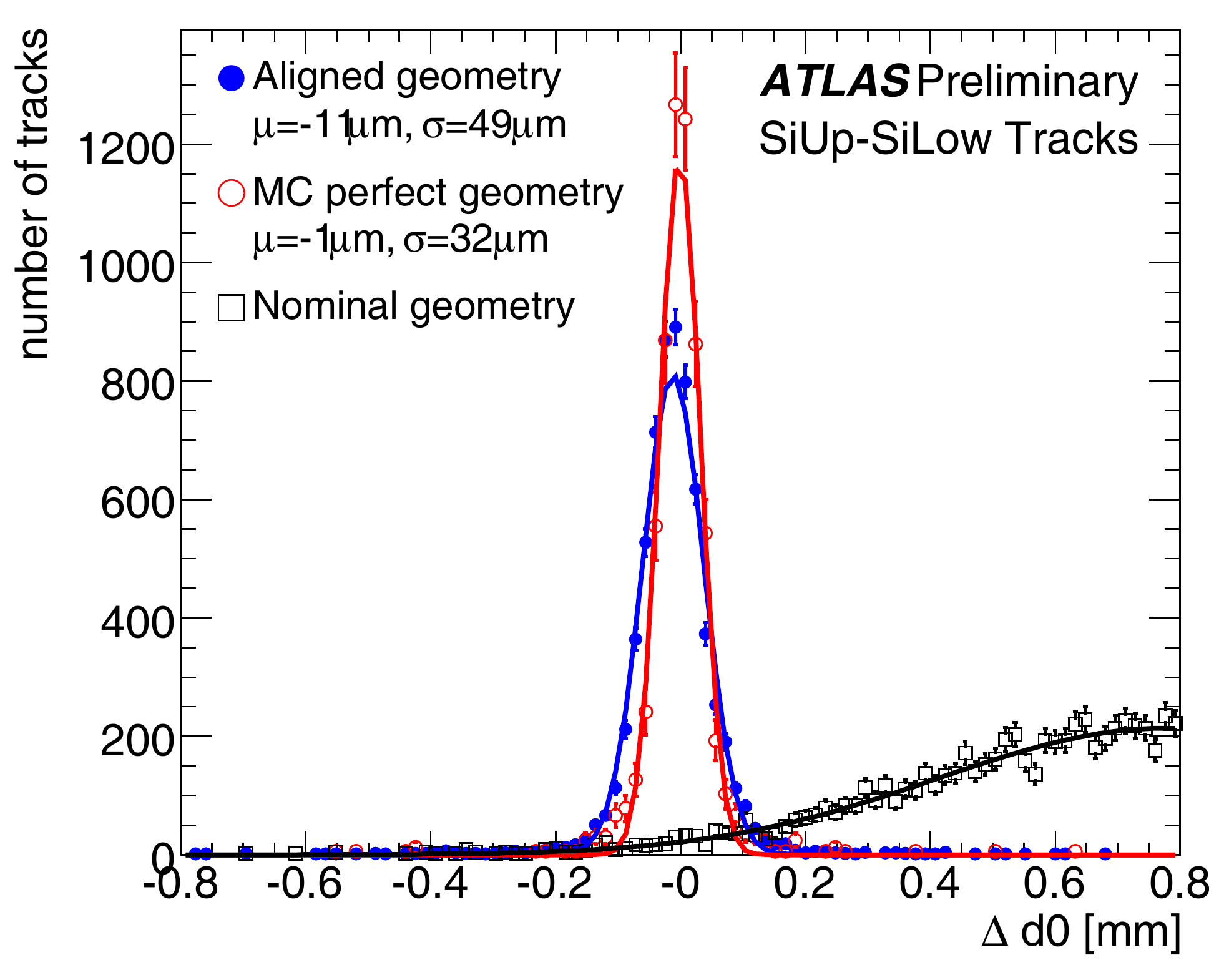}
\caption[]{The difference in the $d_0$ track parameter between the two split tracks in the inner detector.}
\label{Approved_DeltaD0}
\end{figure}

The quality of the alignment of the Inner Detector (ID) is illustrated by 
Fig. \ref{Approved_DeltaD0}. Cosmic tracks crossing the entire ID leave hits in both the upper and lower halves of the ID. These tracks can be split near the interaction point and fitted separately, resulting in two collision-like tracks that can then be compared. The plots show the difference in $d_0$, 
the transverse distance of closest approach to the beam axis. Tracks are selected to have $p_T > 2$ GeV, $|d_0|<50$ mm, $|z_0|<400$ mm (in other words they are required to go through the innermost pixel layer). Tracks also are required to have a hit in the Pixel B layer, 3 Pixel hits and in total 7 Silicon hits. 

\subsection*{ATLAS calorimeters}

\begin{figure}[ht]
\centering
\includegraphics[width=1.0\textwidth]{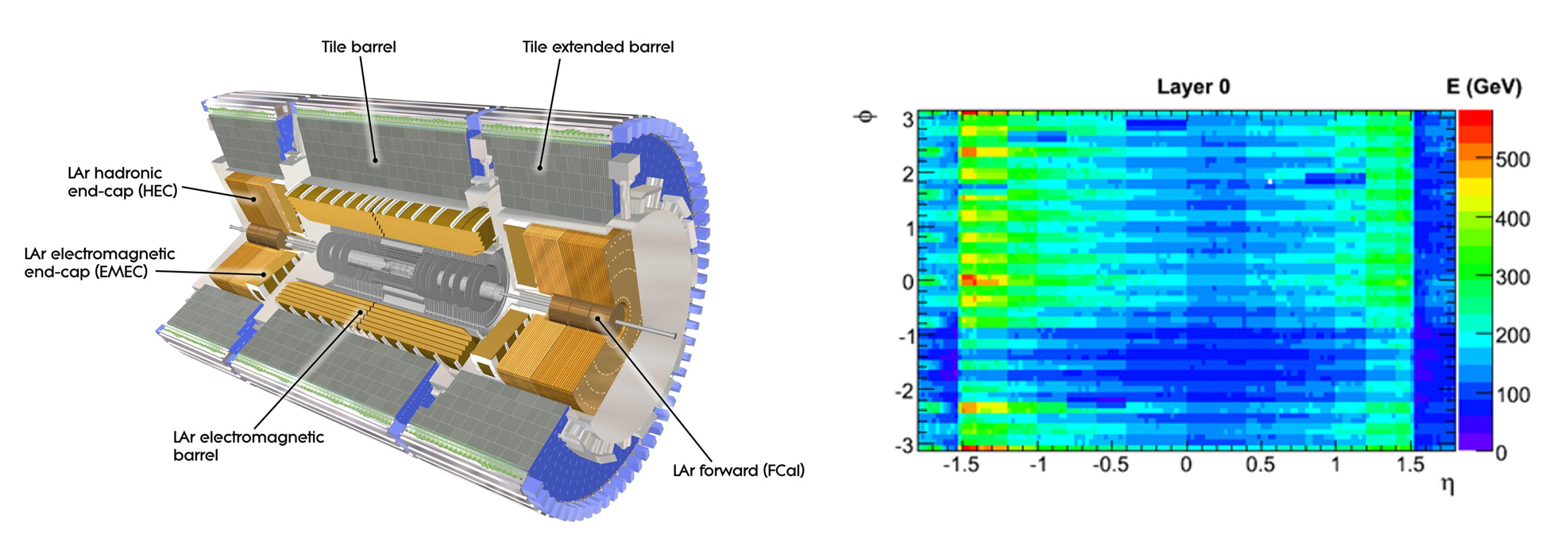}
\caption[]{{\bf Left panel:} The ATLAS calorimeters. {\bf Right panel:} Energy summed over 100 splash events in EM Presampler.}
\label{calo1}
\end{figure}

\begin{figure}[ht]
\centering
\includegraphics[width=1.0\textwidth]{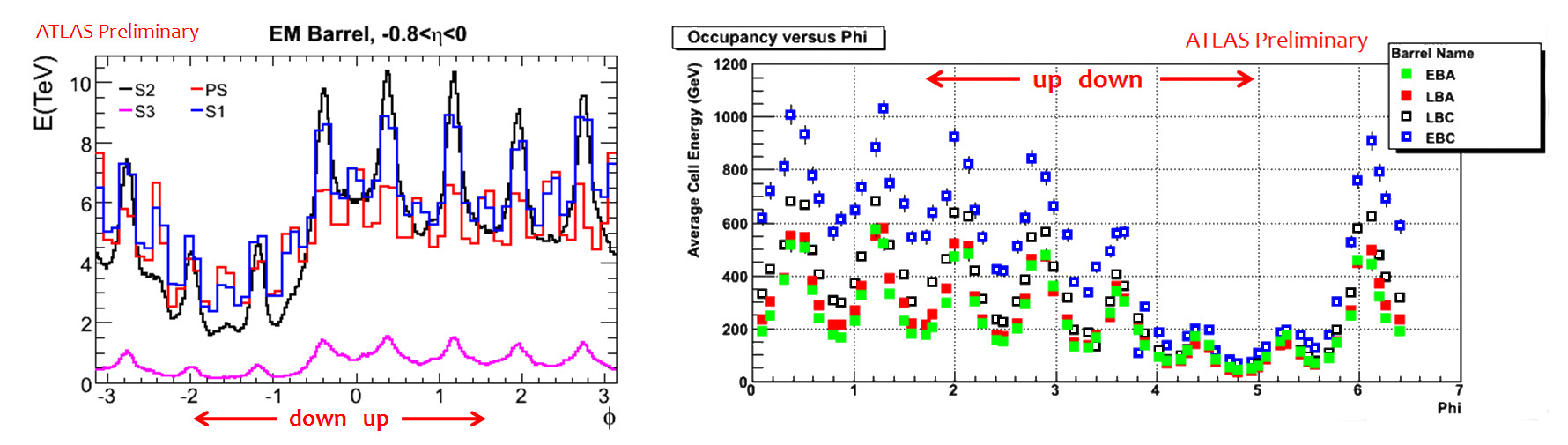}
\caption[]{The sum of 100 splash events from Sept. 10th. {\bf Left panel:} The 8-fold $\phi$ structure of energy measured by the barrel EM calorimeter induced by the toroid endcap is clearly visible at large radius (layers S1, S2, S3). The 16-fold structure is due to additional matter and shielding at low radius (Presampler PS).
{\bf Right panel:} The signal from the tile hadronic calorimeter, again with the same $\phi$ structure. The up-down asymmetry is also due to the material in front of the detector.}
\label{calo2}
\end{figure}

The liquid argon calorimeter operated well in 2008 with 97\% of power supplies operational, 99.92\% channels of the calorimeter were fully working. The tasks for 2009 include repairs of faulty power supplies and improvements of the monitoring. 
The performance of the barrel EM (ElectroMagnetic) liquid argon calorimeter, in particular the presampler layer, is shown in the right panel of Fig. \ref{calo1} and on the left panel of Fig. \ref{calo2}. Higher energy is observed in the upper hemisphere ($\phi$ positive), reflecting the origin of the "splash" particles and due to the location of the tunnel and detector.

The tile calorimeter also performed very well with 99.2\% of power supplies and controls working within specifications and 98.6\% of the cells giving signal. The calibration is ready or being commissioned, cell equalization reached a level of few percent. 
The tasks for 2009 are 
further tuning of the $^{137}$Cs cell inter-calibration, monitoring of PMT stability, data quality monitoring. The right panel of Fig. \ref{calo2} illustrates the performance of the Tile calorimeter in splash events.

\subsection*{Muons}
The ATLAS muon system with its huge air core toroids and all subdetectors performed also well during 2008 with further improvement expected for 2009 (e.g. improvement from 98.3\% monitored drift tubes channels already working to get 99.8\% operational after the shutdown). 

\begin{figure}[ht]
\centering
\includegraphics[width=0.9\textwidth]{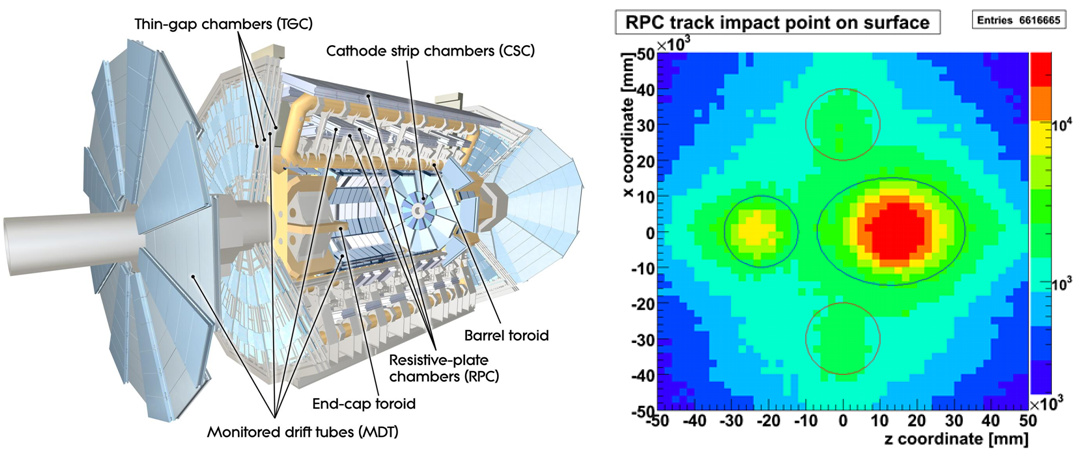}
\caption[]{{\bf Left panel:} The ATLAS muon detectors: Monitored Drift Tubes (MDT) and Cathode Strip Chambers (CSC) for precision tracking; Thin-Gap Chambers (TGC) and Resistive-Plate Chambers (RPC) for trigger. {\bf Right panel}: Cosmic muon map reconstructed by off-line RPC standalone muon monitoring projected on surface (81 m above the nominal interaction point).}
\label{muons}
\end{figure}

The performance of the muon detectors is illustrated by the right panel of Fig. \ref{muons}, which shows the density of recorded cosmic muons projected backwards to the surface. These muons map the ATLAS cavern and the access shafts, which they met on their flight into ATLAS.

\subsection*{Forward detectors}
In addition to the already mentioned central subdetectors three forward detectors are being built: LUCID (Luminosity Cerenkov Integrating Detector), which should serve as the main relative luminosity monitor in ATLAS, 
ZDC (Zero Degree Calorimeter), which is designed to detect spectator neutrons in heavy ion collisions and ALFA (Absolute Luminosity for ATLAS) for absolute luminosity measurement.  
LUCID is a running detector in advanced status of commissioning; nevertheless, some consolidation and repairs are ongoing during the shutdown period. ZDC is at an advanced state of construction, ALFA is expected to be ready in 2010.
  
\begin{figure}[ht]
\centering
\includegraphics[width=1.0\textwidth]{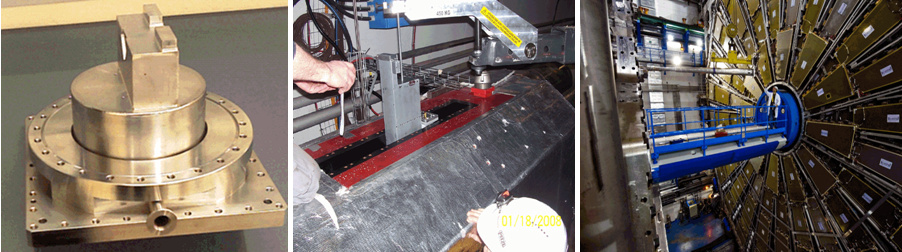}
\caption[]{{\bf Left panel:} ALFA (Absolute Luminosity for ATLAS) at 240 m. {\bf Middle panel:} ZDC (Zero Degree Calorimeter ) at 140 m. {\bf Right panel:} LUCID (Luminosity Cerenkov Integrating Detector) at 17 m from the collision point.}
\label{forward_detectors}
\end{figure}

\subsection*{Data acquisition}

\begin{figure}[ht]
\centering
\includegraphics[width=0.6\textwidth]{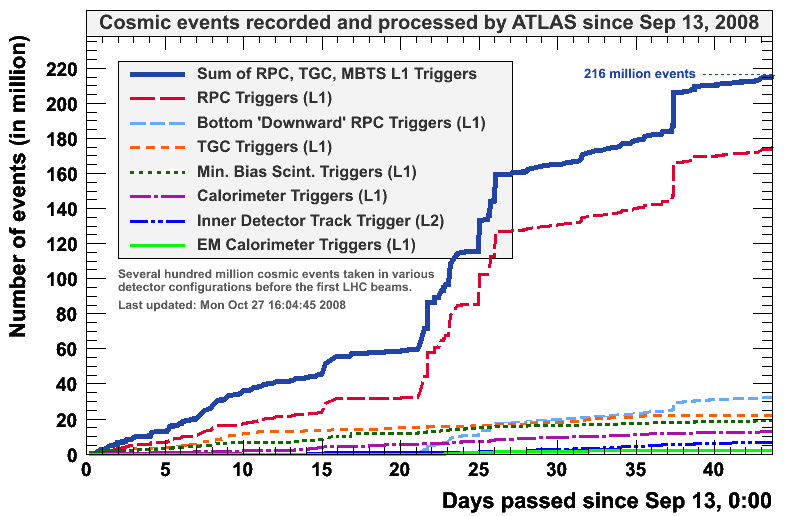}
\caption[]{Cosmic data since Sept 13, 2008. 216 M events. 400,000 files in 21 inclusive streams. }
\label{image13}
\end{figure}

The performance of the data acquisition chain is briefly illustrated by Fig. \ref{image13}, which shows the number of cosmic events recorded with different triggers. The results presented here (and in the previous sections) can be found on the ATLAS public web site \cite{TWiki}.

\newpage
\section{Heavy Ion Physics with the ATLAS Detector at the LHC}
\begin{figure}[ht]
\centering
\includegraphics[width=0.6\textwidth]{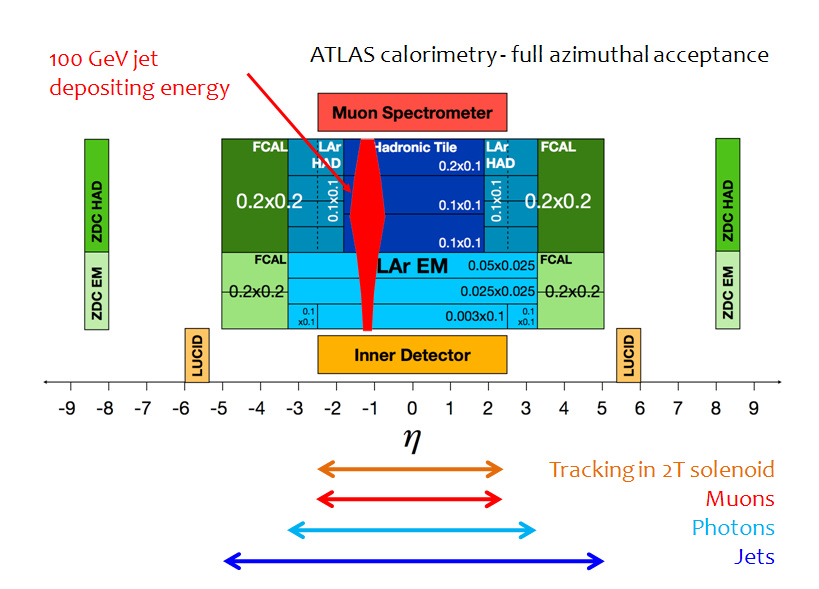}
\caption[]{Acceptance of ATLAS.}
\label{acceptance}
\end{figure}

This short part will show only briefly some results from the forthcoming Physics Performance Report devoted to heavy ion physics. 
Although ATLAS was not originally designed for use in heavy ion collisions, its unprecedented acceptance (see Fig. \ref{acceptance}) and other properties make it an excellent tool to study heavy ion collisions. 
More details are presented in parallel contributions to this conference: Jets in \cite{grau}, Quarkonia and Z$^0$ in \cite{rosati} and Direct photons in \cite{baker}. 

\begin{figure}[ht]
\centering
\includegraphics[width=0.9\textwidth]{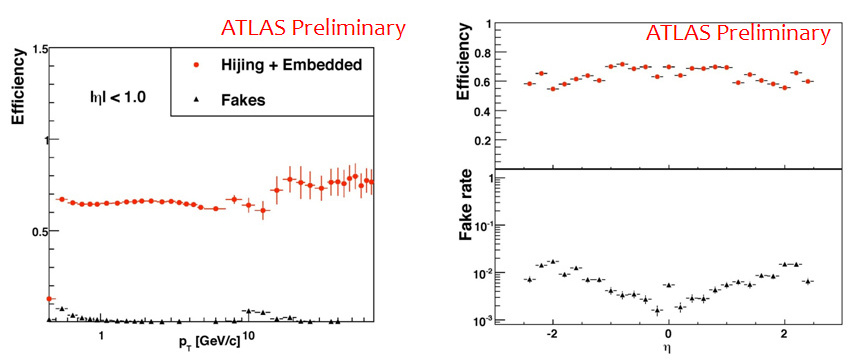}
\caption[]{{\bf Left panel:} Tracking efficiency and fake rate in $|\eta| < 1$ extracted from a sample of central ($b=2$ fm, d$N/$d$\eta=2700$) HIJING \cite{hijing} events produced with quenching effects turned off. {\bf Right panel:} Top: Tracking efficiency as a function of pseudorapidity for tracks with $3 < p_T < 8$ GeV extracted from the same central sample of events. Bottom: Fake rate as function of pseudo-rapidity for the same tracks as above.}
\label{tracking}
\end{figure}

The tracking performance is illustrated in Fig. \ref{tracking}. The efficiency of about 70\% is sufficient for physics studies, and is being optimized, and the fake rate above 1 GeV/c is negligible, and nearly independent of pseudorapidity.

\begin{figure}[ht]
\centering
\includegraphics[width=1.0\textwidth]{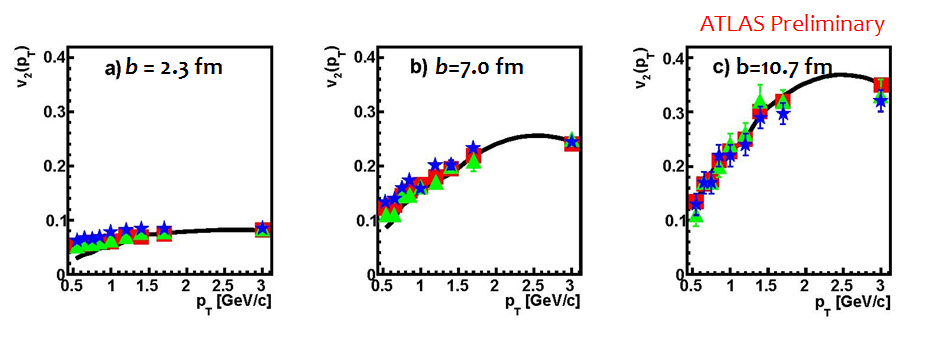}
\caption[]{Transverse momentum dependence of the reconstructed $v_2$: 
from the event plane method (red squares), 
two-particle correlations (blue stars), 
the Lee-Yang Zeros method (green triangles)
input flow as extrapolated from RHIC data (line)}
\label{flow}
\end{figure}

\newpage
Thanks to the complete azimuthal coverage we can well reproduce the simulated elliptic flow, see Fig. \ref{flow}. The use of different methods offer sensitivity to non-flow effects. 

\begin{figure}[ht]
\centering
\includegraphics[width=1.0\textwidth]{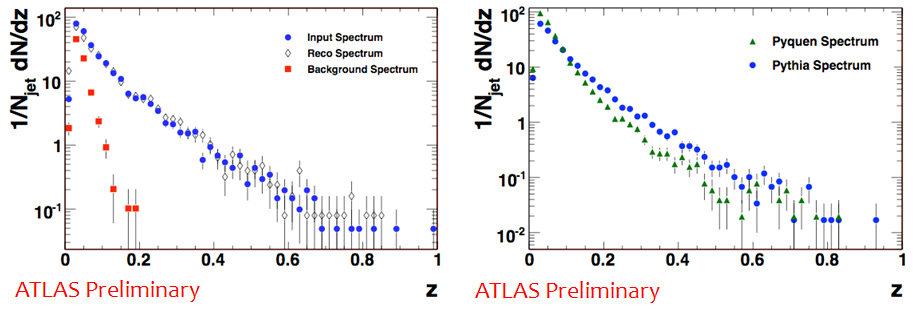}
\caption[]{{\bf Left panel:} Reliable reconstruction of fragmentation function $D(z)$ ($z$ is the longitudinal momentum fraction of a jet carried by a fragment):
Reconstructed tracks with $p_T > 2$ GeV matching calorimeter jets. {\bf Right panel:} The scale of possible modifications of fragmentation function by quenching – comparison of PYTHIA \cite{pythia} and PYQUEN \cite{pyquen}}
\label{jets}
\end{figure}

The sophisticated longitudinally-segmented calorimetry of the ATLAS detector, together with methods for subtracting the background from the underlying event, will allow the study of jets and their modification in the strongly-coupled medium. An illustration of this is shown in Fig. \ref{jets}. 

\begin{figure}[ht]
\centering
\includegraphics[width=0.6\textwidth]{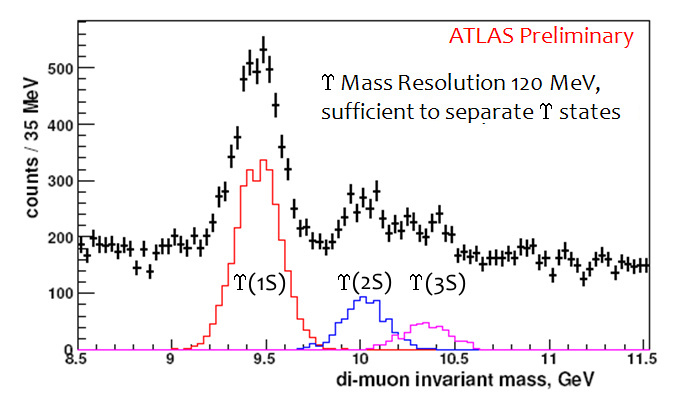}
\caption[]{Di-muon invariant mass distribution as expected for one month of data, taking into account acceptance and efficiency, for decay muons in the barrel region only ($|\eta| < 1$).}
\label{quarkonia}
\end{figure}

Di-muon measurements with the invariant mass resolution good enough to separate the different upsilon states (as shown in Fig. \ref{quarkonia}) can elucidate the details of quarkonia suppression.

\newpage
\section{Summary}

\begin{itemize}
\item ATLAS is fully operational, recorded several hundred million cosmic events.
\item Ongoing activities enable further detector improvements, calibration, refinement of monitoring, software tools.
\item Extensive preparations for Pb+Pb program show a promising performance of ATLAS for heavy ion beams.  
\item ATLAS heavy-ion group will participate in initial p+p data taking to get reference data for heavy ion program and to tune the analysis techniques.
\end{itemize}



\end{document}